\documentclass[12pt]{article}
\pdfoutput=1
\catcode`\@=11
\@addtoreset{equation}{section}

\global\arraycolsep=2pt 
\usepackage{mathrsfs, amsbsy, amssymb, latexsym, amsfonts, amscd, amsmath, cite} 
\usepackage{graphicx,color}

\newcommand{\Tr}{{\rm Tr}}

\allowdisplaybreaks

\begin{document}

\begin{flushright}
\parbox{4cm}
{KUNS-2546}
\end{flushright}

\vspace*{1.5cm}

\begin{center}
{\Large\bf Chaos in the BMN matrix model}
\vspace*{2.5cm}\\
{\large Yuhma Asano$^{\ast}$\footnote{E-mail:~yuhma@gauge.scphys.kyoto-u.ac.jp}, 
Daisuke Kawai$^{\ast}$\footnote{E-mail:~daisuke@gauge.scphys.kyoto-u.ac.jp} 
and Kentaroh Yoshida$^{\ast}$\footnote{E-mail:~kyoshida@gauge.scphys.kyoto-u.ac.jp}} 
\end{center}
\vspace*{0.25cm}
\begin{center}
$^{\ast}${\it Department of Physics, Kyoto University \\ 
Kyoto 606-8502, Japan.} 
\vspace*{0.25cm}
\end{center}
\vspace{1cm}

\begin{abstract}
We study classical chaotic motions in the Berenstein-Maldacena-Nastase (BMN) matrix model. 
For this purpose, it is convenient to focus upon a reduced system composed of two-coupled 
anharmonic oscillators by supposing an ansatz. 
We examine three ans\"atze:   
1) two pulsating fuzzy spheres, 2) a single Coulomb-type potential, and 3) integrable fuzzy spheres. 
For the first two cases, we show the existence of chaos 
by computing Poincar\'e sections and a Lyapunov spectrum. 
The third case leads to an integrable system. 
As a result, the BMN matrix model is not integrable 
in the sense of Liouville, though there may be some integrable subsectors. 
\end{abstract}

\setcounter{footnote}{0}
\setcounter{page}{0}
\thispagestyle{empty}

\newpage

\tableofcontents

\section{Introduction}

One encounters non-linear differential equations everywhere in theoretical physics. 
String theory is no exception, as a matter of course. 
For example, when a string is moving on a curved space, 
the classical equation of motion becomes non-linear generally. 
If the curved background has a special coset-structure like AdS$_5\times$S$^5$\,, 
the string theory is classically integrable in the sense that the Lax pair exists \cite{BPR}. 
Then classical solutions can be studied analytically, for example, 
by using the classical inverse scattering method. 
However, in most cases, one cannot solve analytically non-linear differential equations. 
The systems are non-integrable and exhibit chaos. 

\medskip 

Recently, chaotic string solutions have been well studied in various ways 
\cite{AdS-soliton,T11,D-brane,NI,NR}. 
Motivation lies on potential applications to the AdS/CFT correspondence \cite{M,GKP,W} 
beyond integrability \cite{review}, 
but it has not been clarified yet what corresponds to chaotic strings in the gauge-theory side. 
Note that chaotic motions cannot be described 
with analytic functions. This is a characteristic of chaos. 
So it may be intriguing to ask whether the correspondence holds even in the non-analytic region or not. 
One can test it at a deeper level.

\medskip 

Along this direction, one may focus upon the M-theory dynamics. 
A matrix model description of light-front M-theory is proposed 
by Banks-Fischler-Shenker-Susskind \cite{BFSS}. 
It is called the BFSS matrix model. Chaos in this model 
was studied in \cite{chaos-BFSS} by following the way in classical Yang-Mills (YM) theories \cite{YM}. 
However, there are some unsatisfactory points. 
In the BFSS matrix model, there are flat directions which may cause unbounded motions. 
Then the motions may conflict with the boundedness of chaos during long-time numerical computation 
of classical trajectories. 
Another point is a technical problem. In \cite{chaos-BFSS}, the largest Lyapunov exponents  
were computed by reducing the system to simple models. 
However, it does not seem that the results converge on definite values. 
In addition, the time scale of (would-be) convergence is also much larger 
than the inverse of the Lyapunov exponent. 
Thus, it would be worth revisiting chaotic motions in the BFSS matrix model. 

\medskip 

A more appropriate system is the Berenstein-Maldacena-Nastase (BMN) 
matrix model \cite{BMN}. It contains mass terms\footnote{
For chaotic motions in a YM-Higgs system, see \cite{deformed-YM}.} 
and flat directions are lifted up. Hence there is no subtlety with the boundedness 
of chaos. In addition, the Myers term may give rise to a separatrix 
structure in the phase space. It means a bifurcation of the behavior of 
classical solutions and may be a source of chaos.
In total, the BMN matrix model is a good laboratory 
to study chaotic motions. 

\medskip 

In this paper, we will study classical chaos in the BMN matrix model. 
For this purpose, it is useful to reduce the full system 
to a simple model composed of two-coupled 
anharmonic oscillators by supposing an ansatz. We examine three cases:   
1) two pulsating fuzzy spheres, 2) a single Coulomb-type potential, and 3) integrable fuzzy spheres. 
For the first two cases, we show the existence of chaos 
by computing Poincar\'e sections and a Lyapunov spectrum. 
The third case leads to an integrable system. 
As a result, the BMN matrix model is not integrable 
in the sense of Liouville, though there may be some integrable subsectors. 

\medskip 

This paper is organized as follows. In section 2 we give a brief review 
of chaos. In particular, the characteristics of chaos in the Hamiltonian system. 
Section 3 considers the H\'enon-Heiles system as an example of non-integrable 
Hamiltonian systems. 
In section 4 we consider chaotic motions and an integrable subsector 
in the BMN matrix model. 
Section 5 is devoted to conclusion and discussion. 
In Appendix A, the computation scheme of Lyapunov spectrum is summarized.

\section{What is chaos?}

Probably, anybody has said the term, ``chaos''.  
For example, when you see the top of the desk in a mess, 
you would say ``It's chaos''. So it seems likely that ``chaos'' implies 
that some things are completely messed up. 
In fact, this point captures an aspect of chaos, {\it randomness}. 

\medskip 

There would be no universally accepted definition of chaos. 
The following is a list of characteristics of chaos commonly seen in almost any books:  
\begin{enumerate}
\item High sensitivity to initial conditions, 
\item Randomness of motions (continuous spectrum),
\item Non-periodic and bounded motions (non-analyticity).
\end{enumerate}

The item 1) is significant and it is sometimes called 
{\it the Butterfly effect}. One may know the following famous sentence, 
``Does the flap of a butterfly's wings in Brazil set off a tornado in Texas?'' 
Here the flapping wing corresponds to a small change of the initial condition 
and it may cause a chain of events leading to large scale phenomena. 

\medskip 

Suppose a continuous dynamical system described by a set of 
non-linear differential equations.  
The system is deterministic because the dynamics is determined by 
fixing an initial condition.  
But one cannot predict the motions generally after long time has passed because of the non-linearity. 
That is, it is impossible to forecast weather in the far future. 
This indicates that some information is lost dynamically 
and it is measured by the Kolmogorov-Sinai (KS) entropy. 
This entropy production is concerned with the randomness [the item 2)].       

\medskip 

The item 3) can be intuitively understood. If the motion is periodic, then 
it does not satisfy the other properties. It exhibits the line spectrum 
and no sharp sensitivity to initial conditions. 
To see that the boundedness is necessary, let us consider a simple system $\dot x=x$\,. 
The orbits run away exponentially to infinity, hence 
it seems to exhibit high sensitivity to initial conditions. 
However, it just means an attracting fixed-point at infinity 
and does not indicate chaotic motions. The recurrence 
of motions should be required implicitly so as to exclude fixed points.    
Non-analyticity follows from the continuous spectrum. 
The smooth behavior in the Fourier space indicates a random, non-analytic motion 
in the real space (i.e., the motion is continuous but 
not differentiable everywhere). Hence chaotic motions do not behave in an analytic way 
in comparison to (quasi-)periodic ones.
Indeed, if the motion is described by an analytic function, then one can predict 
the motion for an arbitrarily long time and it does not satisfy the item 1).  

\medskip 

There are three methods to see chaos, 1) Poincar\'e sections, 
2) Lyapunov spectrum and 3) power spectrum. 
Since some readers are not familiar to these issues, 
it would be helpful to introduce each of them below.

\subsection{Poincar\'e sections}

To study complex trajectories in the phase space, 
it is helpful to use a Poincar\'e section, 
which is a surface of section intersecting the flow in the phase space.
It is defined so that all of the possible orbits cross the section repeatedly.
The dimension of the flow (i.e., the dynamical system) is 
the number of the first-order ordinary differential equations\footnote{
Note that, in an autonomous Hamiltonian system, 
all of the orbits are restricted onto a lower dimensional manifold 
due to the constancy of the energy.
In this sense, the dimension of this dynamical system is 
one lower than twice the dimension of real space.
In general, the dimension of a dynamical system is 
lower than twice the dimension of real space   
by the number of extra conserved charges.
},
and the dimension of Poincar\'e sections is one lower than it. 
On a Poincar\'e section, an orbit is mapped to dots, 
which are the intersection points of the orbit and the Poincar\'e section.
Since the dynamics is governed by a set of differential equations,
the original trajectory can be reconstructed from the dots on the Poincar\'e section.
In this sense, the flow and the Poincar\'e section are equivalent, 
while there is a degree of freedom to choose the surface in practice. 
It is the usual to pick up only the dots at which the trajectory intersects 
from only the one side to the other side.
This constraint allows any dots on a Poincar\'e section to uniquely determine 
the next point of the trajectory in autonomous Hamiltonian systems.
Suppose a Hamiltonian system with a four-dimensional phase space $(q_1,q_2;p_1,p_2)$\,. 
A Poincar\'e section is a two-dimensional surface, for example, with $q_2=0$ and $p_2 >0$\,.  

\medskip 

By changing initial values, a set of trajectories is generated 
and it contains several kinds of motions. 
The Poincar\'e section is helpful to figure out their behaviors.
Periodic orbits are projected onto the Poincar\'e section 
as finite numbers of dots on closed contours.
A periodic trajectory comes back to a point on the Poincar\'e section
after its period. Quasi-periodic orbits are also projected as dots on closed contours,
but the resulting portrait becomes dense. 
They are understood as non-resonant tori.
Completely chaotic orbits result in randomly scattered dots on the Poincar\'e section.
These behaviors depend on initial values and generally chaotic motions coexist with 
(quasi-)periodic motions on the section.

\medskip 

Suppose that a small perturbation makes an integrable system be non-integrable. 
Then the Kolmogorov-Arnold-Moser (KAM) theory \cite{Ko,Ar,Mo} claims that 
trajectories incline to be chaotic gradually as the perturbation becomes larger.
When the perturbation is small enough, most of the orbits are (quasi-)periodic. 
These are called KAM tori, which are the remnants of Liouville tori 
in the unperturbed integrable system. In other words, 
almost all of the Liouville tori survive the small perturbation, 
though the tori are distorted and translated.  
As the perturbation becomes larger, chaotic trajectories appear and coexist with KAM tori. 
Then the region of chaotic trajectories enlarges 
while the KAM tori tend to get broken. 
When the perturbation is larger, KAM islands and islets are in the sea of chaos.

\subsection{Lyapunov spectrum}

The next is to introduce the Lyapunov spectrum. 
It is very useful to measure the dynamical information loss. 

\medskip 

Let us first introduce a Lyapunov exponent. 
We basically follow the description in \cite{Strogatz}. 
Suppose that a solution ${\bf x}(t)$ is chaotic 
and consider a nearby point ${\bf x}(t)+\delta(t)$\,. 
Then the deviation $\delta(t)$ grows exponentially 
like 
\begin{eqnarray}
||\delta(t) || \sim ||\delta(0)||\,{\rm e}^{\lambda t}\,.
\end{eqnarray}
Here $\lambda$ is a positive constant called a Lyapunov exponent. 
This behavior indicates that the deviation of the initial value $\delta(0)$ 
is amplified exponentially even if it is very tiny. This is nothing but 
a realization of sensitive dependence on initial conditions. 

\medskip 

In general, the phase space is higher-dimensional. Then 
there are $N$ different Lyapunov exponents for an $N$-dimensional phase space. 
An intuitive explanation is the following. 
Consider time evolution of an infinitesimal sphere of perturbed initial configurations. 
During its evolution, the sphere tends to be distorted into an infinitesimal ellipsoid. 
Let $\delta_k(t)~(k=1,\ldots,N)$ denote the length of the $k$-th principal axis of the ellipsoid. 
Then the deviations behave as 
\[
\delta_k(t) \sim \delta_k(0)\,{\rm e}^{\lambda_k t}\,, \qquad \lambda_k~:\mbox{Lyapunov exponents}. 
\]
The set of the exponents $\lambda_k$ generates a spectrum called the Lyapunov spectrum. 
For large $t$\,, the diameter of the ellipsoid is controlled by the most positive $\lambda_k$\,, 
and it is called the largest Lyapunov exponent. 

\medskip

The Lyapunov spectrum is also related to the dynamical entropy production. 
For example, the sum of all of the positive Lyapunov exponents is identified 
with the KS entropy 
when the system is ergodic. It would be very interesting to look for an application of the KS entropy 
in the context of string theory and M-theory.

\subsection{Power spectrum}

The power spectrum is defined as 
\begin{eqnarray}
I (\omega) \equiv \lim_{\tau\to\infty}\frac{\tau}{2\pi}\left\langle 
\left|
\frac{1}{\tau}\int_0^{\tau}\!dt\,y(t)\,{\rm e}^{-i\omega t}
\right|^2\right\rangle\,,  
\end{eqnarray}
where $\langle y(t) y(s)\rangle$ is a two time correlation function.  
It is quite helpful when one needs to distinguish quasi-periodic motions 
and chaotic motions. When the ratio of periods is irrational, the former leads to dense orbits   
in the phase space 
(the denseness). On the other hand, the latter is related to a stochastic motion (the randomness). 
In the Hamiltonian system, 
there is no attractor because the phase space volume has to be preserved due to Liouville's theorem. 
Hence it is often difficult to distinguish them 
at a first glance. However, the power spectrum shows clear distinction. 
For quasi-periodic motions, the spectrum contains sharp peaks associated with periods 
(the line spectrum)\,. On the other hand, the continuous spectrum 
appears for chaotic motions. 

\medskip 

In the following, we will focus upon the first two methods, Poincar\'e sections 
and a Lyapunov spectrum.

\section{Warming up -- the H\'enon-Heiles system}

In this section, let us consider the H\'enon-Heiles system \cite{HH}. 
This is a famous example of non-integrable Hamiltonian systems containing chaotic motions. 
This model was originally proposed by H\'enon and Heiles (1964) in the study of celestial mechanics.

\subsection*{H\'enon-Heiles system} 

The Hamiltonian is given by 
\begin{eqnarray}
H = \frac{1}{2}(p_1^2 + p_2^2) + \frac{1}{2}(q_1^2+q_2^2) 
+ a \left(q_1^2 q_2 -\frac{1}{3}q_2^3 \right)\,, \label{HH}
\end{eqnarray}
where $a$ is an arbitrary real constant parameter. The system is composed of two 
particles with the same mass (normalized to 1) 
and the interaction is third-order. 
The positions of the particles are described by $q_1$ and $q_2$\,, and the conjugate momenta are $p_1$ and $p_2$\,. 
It is well-known that this system is non-integrable. 

\medskip 

Before going to the detailed analysis, it may be interesting to see that the form 
of the interaction term in (\ref{HH}) is very important. If the interaction term is 
slightly modified like
\begin{eqnarray}
 a \left(q_1^2 q_2 - \frac{1}{3}q_2^3 \right) \qquad 
\longrightarrow \qquad 
 a \left(q_1^2 q_2 + \frac{1}{3}q_2^3 \right)\,, 
 \nonumber 
\end{eqnarray}
then the system becomes classically integrable. The resulting system is often called 
``anti H\'enon-Heiles system''. 

\medskip 

The system with (\ref{HH}) is equivalent to a periodic lattice with three particles, 
\begin{eqnarray}
H &=& H_0 + \frac{1}{3}\alpha \left[(Q_1-Q_2)^3 + (Q_2-Q_3)^3 + (Q_3-Q_1)^3 \right]\,, \nonumber \\ 
H_0 &=& \frac{1}{2}(P_1^2+P_2^2+P_3^2) 
+ \frac{1}{2}\left[(Q_1-Q_2)^2 + (Q_2-Q_3)^2 + (Q_3-Q_1)^2\right]\,. \label{three} 
\end{eqnarray}
The unperturbed Hamiltonian $H_0$ describes a system of three connected harmonic oscillators. 
Hence the equations of motion for $H_0$ are exactly solved by using the normal coordinates. 
By using the normal coordinate for $H_0$\,, one can rewrite the total Hamiltonian $H$\,. 
Then, after dropping the total momentum, the H\'enon-Heiles system is reproduced. 

\medskip 

Here one may sum up higher-order interaction terms with appropriate 
numerical coefficients and obtain an exponential-type potential. 
The resulting system is nothing but a Toda lattice system \cite{Toda}. Hence, inversely speaking, 
the H\'enon-Heiles system may be regarded as a truncation of the Toda lattice system.

\subsection*{Poincar\'e sections and Lyapunov spectrum}

\begin{figure}[htb]
\begin{tabular}{cc}
\includegraphics[scale=.3,angle=-90]{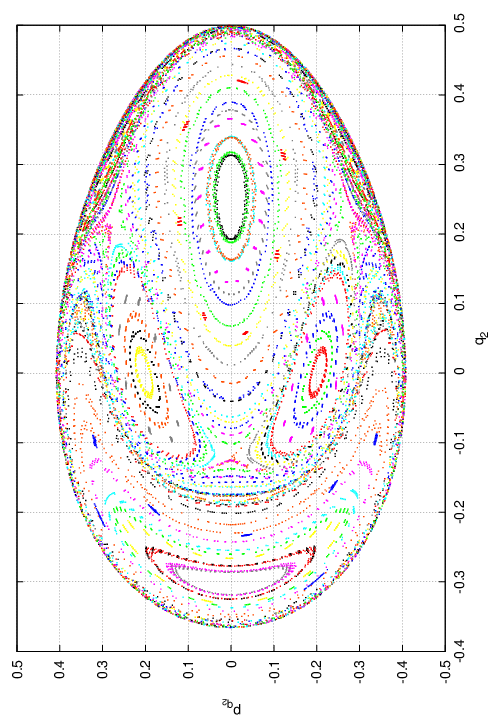} & 
\includegraphics[scale=.3,angle=-90]{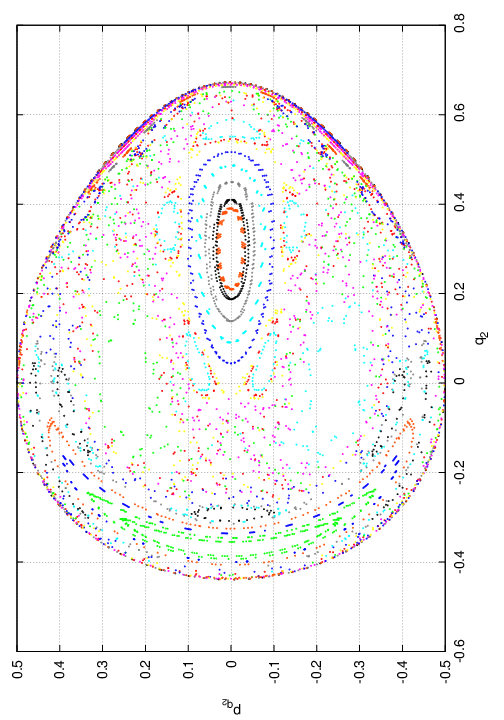} \\ 
{\footnotesize (a) \quad Poincar\'e section with $E=0.0833$} & 
{\footnotesize (b) \quad Poincar\'e section with $E=0.1250$} \\ 
\includegraphics[scale=.3,angle=-90]{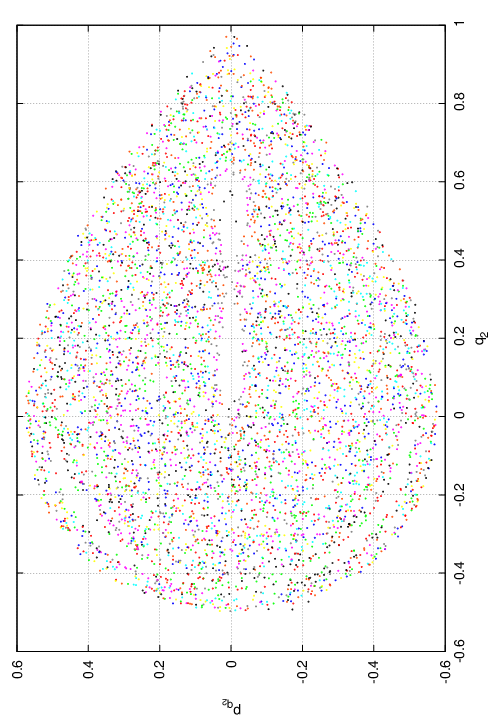} & 
\includegraphics[scale=.31,angle=-90]{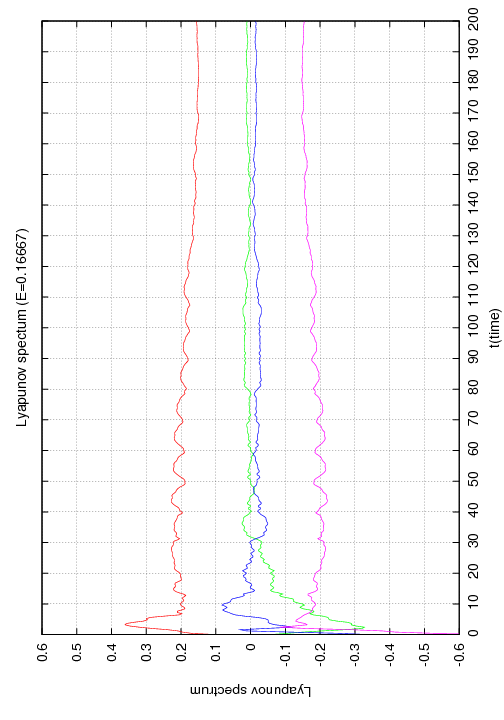} \\ 
{\footnotesize (c) \quad Poincar\'e section with $E=0.16667$} & 
{\footnotesize (d) \quad Lyapunov spectrum with $E=0.16667$}
\end{tabular}
\caption{ \label{Fig1}
\footnotesize Poincar\'e sections and a Lyapunov spectrum.
Different trajectories are indicated by dots with different color, depending on initial conditions.
}
\end{figure}

Let us see chaotic motions in the H\'enon-Heiles system (\ref{HH})\,. 
It is a good exercise to understand chaotic motions in the Hamiltonian system. 
Here we will consider Poincar\'e sections and a Lyapunov spectrum. 
These quantities can be evaluated numerically and the results are 
shown in Fig.\,\ref{Fig1}.

\medskip 

Figures \ref{Fig1} (a), (b) and (c) are Poincar\'e sections for $E=0.0833$, $0.1250$ and 
$0.16667$\,, respectively. The sections are taken with $q_1=0$ and $\dot{q}_1>0$\,. 
We set that $a=1$\,.
In Fig.\,\ref{Fig1} (a), most of the solutions are (quasi-)periodic (KAM tori). 
In Fig.\,\ref{Fig1} (b), chaotic motions appear but still (quasi-)periodic motions coexist. 
In Fig.\,\ref{Fig1} (c), most of the solutions are chaotic.

\medskip 

Figure \ref{Fig1} (d) is a Lyapunov spectrum with $E=0.16667$\,. 
It is computed by taking an initial condition randomly from the region around $q_1=0.0$, $p_1=0.35$, $q_2=-0.115$, $p_2=-0.443$
with the radius $5.0\times 10^{-3}$\,. 
The largest Lyapunov exponent is $\lambda_1 \sim 0.15$\,. 
This positive value indicates high sensitivity to initial conditions 
and provides a positive support for the existence of chaos in this system. 
As a feature of the Hamiltonian system, the spectrum is symmetric 
and the total sum of all of the exponents becomes zero. The remaining two exponents become zero due to the energy conservation.

\medskip 

The features presented in Fig.\,\ref{Fig1} are characteristics of chaos in the Hamiltonian system. 
In the next section, we will show similar results in the case of the BMN matrix model.

\section{Chaotic motions in the BMN matrix model}

We study here classical chaotic motions in the BMN matrix model. 
It is useful to reduce the full theory to a simple model 
so as to extract a mechanism which induces chaotic motions. 
We examine three reduction ans\"atze and argue whether chaotic motions 
exist or not.

\subsection{Setup}

The BMN matrix model \cite{BMN} was originally proposed as a realization of M-theory 
on the maximally supersymmetric pp-wave background \cite{KG}. 
That is, it can be regarded as a deformation of the BFSS matrix model \cite{BFSS}.
This model can also be derived 
as the matrix regularization of a supermembrane theory on the pp-wave \cite{DSR,SY} 
by following a seminal work \cite{dWHN}. 

\medskip 

The BMN matrix model is a one-dimensional $U(N)$ gauge theory composed of 
matrix-valued variables $X^{r}(t)~(r=1,\ldots,9)$\,, a gauge field $A$ and 16 fermions. 
For the present purpose, we will focus upon the bosonic sector.   

\medskip 

The bosonic part of the classical action is given by 
\begin{eqnarray}
 && S = \int\! dt\, \Tr \Bigg[ 
 \frac{1}{2}(D_t X^r)^2 
 +\frac{1}{4}[X^r,X^s]^2 
 \nonumber \\ && \hspace*{3cm}
 -\frac{1}{2}\left(\frac{\mu}{3}\right)^2X_{i}^2
 -\frac{1}{2}\left(\frac{\mu}{6}\right)^2X_{a}^2
 -\frac{\mu}{3}i\epsilon_{ijk}X^iX^jX^k
 \Bigg]\, ,
\end{eqnarray}
where $i,j=1,2,3$ and $a,b=4,\ldots, 9$\,. The covariant derivative $D_t$ is defined as 
\[
D_t X^r \equiv \dot{X}^r -i [A,X^r]\,. 
\] 
The symbol ``$\cdot$'' denotes the derivative with respect to $t$\,. The real constant parameter $\mu$ 
measures the deformation. When $\mu =0$\,, the system is reduced to the BFSS matrix model. 
When $\mu \neq 0$\,, mass terms and the Myers term \cite{Myers} are turned on\footnote{
Note that the mass ratio for $X^i$ and $X^a$ is fixed by requiring the maximal 32 supersymmetries 
for the pp-wave background \cite{KG,BMN}. 
The coefficient of the Myers term is fixed as well. 
In this sense, supersymmetries are implicitly concerned 
with our analysis, though the present argument is restricted to the bosonic sector. 
}. 

\medskip 

In the BMN matrix model, the classical trajectories are bounded due to the mass terms. 
Hence there is no flat direction in comparison to the BFSS case. 
This characteristic would be so important as to study chaotic behavior definitely 
for sufficiently long time. 

\medskip 

Due to the presence of the Myers term, there exist fuzzy sphere vacua 
as well as the trivial vacuum $X^r =0$\,. 
Thus the BMN matrix model has the rich vacuum structure. 
Furthermore, the Myers term leads to a separatrix structure in the phase space 
and may provide a new source of chaotic motions. 

\medskip

We will work with the Weyl gauge $A=0$ hereafter. 
Then the classical equations of motion are given by 
\begin{eqnarray}
 0 &=& \ddot X^i
 +\sum_{r=1}^9[X^r,[X^r,X^i]]
 +\left(\frac{\mu}{3}\right)^2X^i
 +i\mu \sum_{j,k=1}^3\epsilon^{ijk}X_jX_k \qquad (i=1,2,3) \,, \nonumber \\ 
  0 &=&  \ddot X^a
 + \sum_{r=1}^9[X^r,[X^r,X^a]]
 +\left(\frac{\mu}{6}\right)^2X^a \qquad (a=4,\ldots,9) \,. \label{eom}
\end{eqnarray}
In addition, one has to take account of the Gauss law constraint, 
\begin{eqnarray}
 0 = \sum_{r=1}^9\,[X^r, \dot{X}^r]\,.
 \label{Gauss}
\end{eqnarray}
This comes from the equation of motion of $A$\,. 

\medskip 

In the next subsection, we will study classical solutions of the equations of motion (\ref{eom}) 
equipped with the Gauss law constraint (\ref{Gauss})\,.

\subsection{Reduction to simple models}

To argue classical chaotic motions in the BMN matrix model, 
it is sufficient to study a simple reduced system by imposing an ansatz.  
We will examine three examples of the reduction ansatz below.

\subsubsection*{(1) \quad Two pulsating fuzzy spheres $(N=2)$}

First of all, let us suppose the following $2 \times 2$ ansatz of classical solutions: 
\begin{eqnarray}
&& X^i = r(t)\, \frac{\sigma^i}{2}\,, 
\qquad X^{a'} = x(t)\, \frac{\sigma^{a'-3}}{2} \quad (a'=4,5,6)\,, \nonumber \\ 
&& X^7=X^8=X^9=0\,.
\label{ansatz1}
\end{eqnarray}
Here $\sigma^i~(i=1,2,3)$ are the standard Pauli matrices. Then 
$x(t)$ and $r(t)$ are real functions to be determined. 
This ansatz (\ref{ansatz1}) describes two pulsating fuzzy spheres 
and satisfy the Gauss law constraint (\ref{Gauss}). 

\medskip 

With the ansatz (\ref{ansatz1})\,, the classical equations of motion (\ref{eom}) 
are reduced to 
\begin{align}
 0&=\ddot r
 +\left( \frac{\mu}{3} \right)^2r
 -\mu r^2
 +2r^3
 +2rx^2\,,
 \nonumber \\
 0&=\ddot x
 +\left( \frac{\mu}{6} \right)^2x
 +2x^3
 +2xr^2\,. \label{s1}
\end{align}
The resulting system (\ref{s1}) is a non-linear dynamical 
system in which two anharmonic oscillators are coupled. 
Note that the system (\ref{s1}) can also be obtained from the Lagrangian, 
\begin{align}
 L&=
 \frac{1}{2}\dot r^2
 -\frac{1}{2}r^2\left( r-\frac{\mu}{3} \right)^2
 +\frac{1}{2}\dot x^2
 -\frac{1}{2}\left(\frac{\mu}{6}\right)^2x^2
 -\frac{1}{2}x^4
 -r^2x^2\,. 
\end{align}

\medskip 

\begin{figure}[htbp]
\begin{tabular}{cc}
\includegraphics[scale=.3,angle=-90]{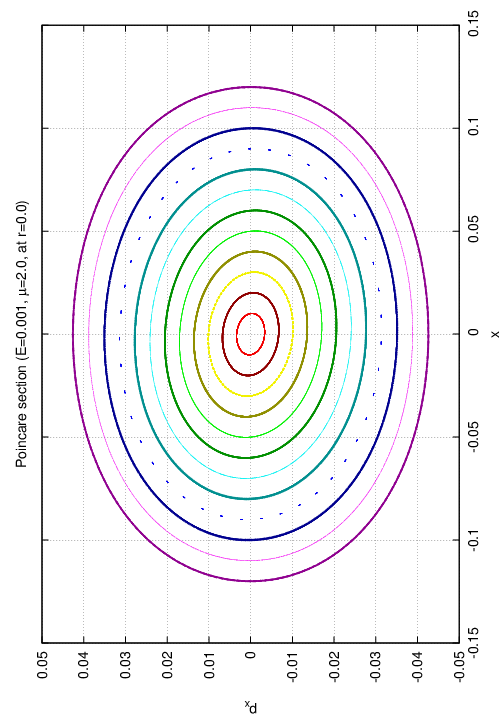} & 
\includegraphics[scale=.3,angle=-90]{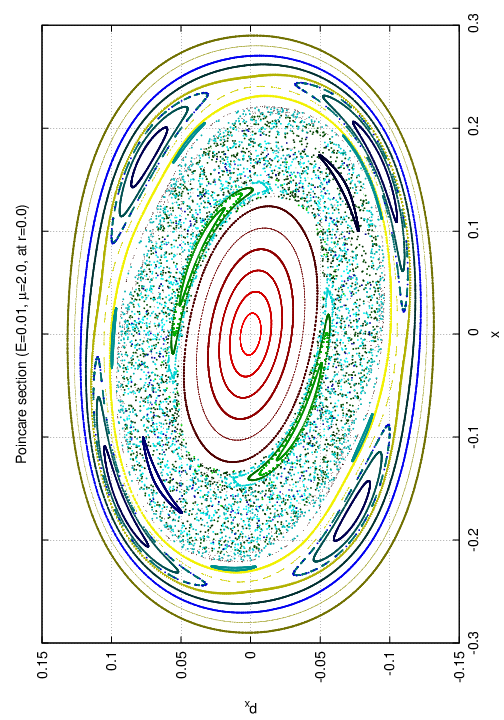} \\
{\footnotesize (a) \quad Poincar\'e section with $E=0.001$} & 
{\footnotesize (b) \quad Poincar\'e section with $E=0.01$} \\
\includegraphics[scale=.3,angle=-90]{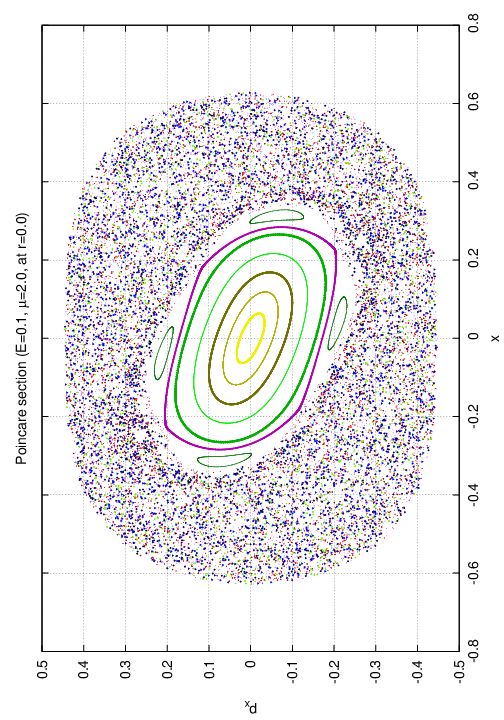} & 
\includegraphics[scale=.3,angle=-90]{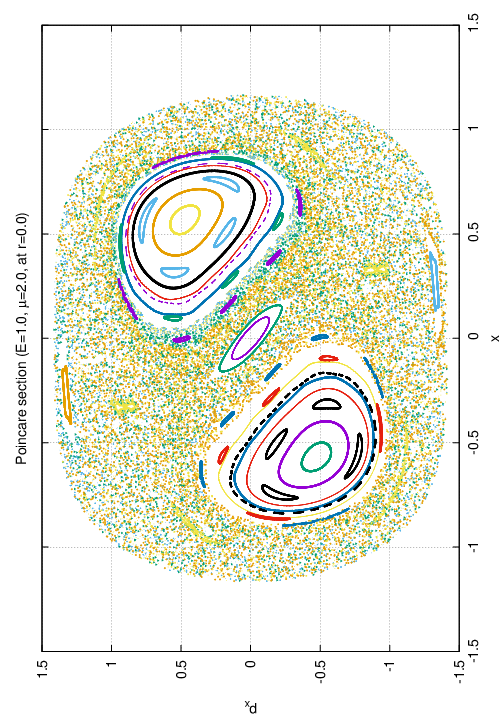} \\ 
{\footnotesize (c) \quad Poincar\'e section with $E=0.1$} &  
{\footnotesize (d) \quad Poincar\'e section with $E=1.0$} \\
\includegraphics[scale=.3,angle=-90]{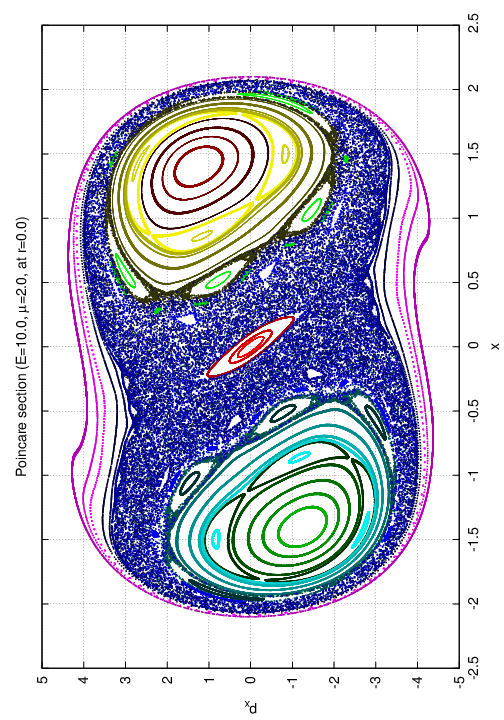} & 
\includegraphics[scale=.3,angle=-90]{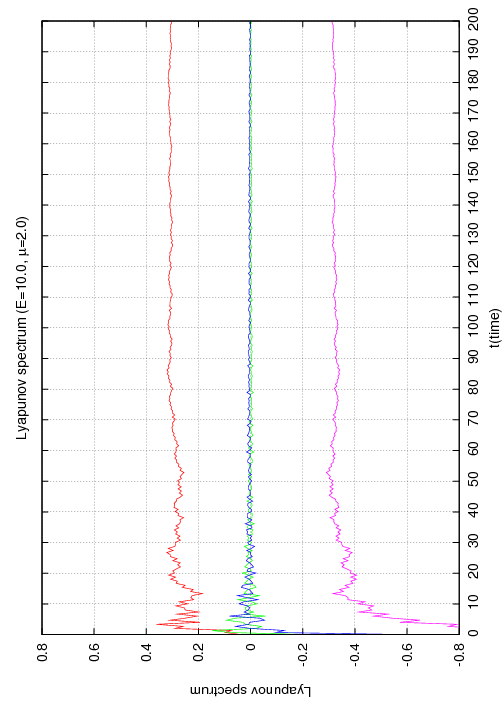} \\ 
{\footnotesize (e) \quad Poincar\'e section with $E=10$} &  
{\footnotesize (f) \quad Lyapunov spectrum with $E=10$} 
\end{tabular}
\caption{ \label{Poincare-BMN1}
\footnotesize Poincar\'e sections and a Lyapunov spectrum 
with the ansatz \eqref{ansatz1}. 
}
\end{figure}

Poincar\'e sections are presented in Figs.\,\ref{Poincare-BMN1} (a)-(e). 
The sections are taken with $r=0$ and $\dot{r} >0$\,. 
For $E=0.001$, 0.01, 0.1,1.0 and 10 with $\mu=2.0$, the sections are numerically computed. 
When $E=0.001$\,, the energy is sufficiently small and only (quasi-)periodic trajectories appear 
[Fig.\,\ref{Poincare-BMN1} (a)].
When $E=0.01$\,, one can see chaotic motions between two layers [Fig.\,\ref{Poincare-BMN1} (b)]. 
This is a typical example of local chaos. 
When $E=0.1$\,,  KAM tori around the outer circle are destroyed and 
a ring of global chaos appears, according to the Chirikov overlap criterion. 
There is a big island in the sea of chaos [Fig.\,\ref{Poincare-BMN1} (c)]. 
When $E=1.0$\,, the island in Fig.\,\ref{Poincare-BMN1} (c) splits into two islands 
[Fig.\,\ref{Poincare-BMN1} (d)]. 
When $E=10$\,, (quasi-)periodic orbits appear outside the sea of chaos 
[Fig.\,\ref{Poincare-BMN1} (e)].  
This is the onset of re-ordered motions. This phenomenon is quite natural 
because motions tend to be harmonic oscillation as the energy becomes much higher.

\medskip 

The Lyapunov spectrum for $E=10$ and $\mu=2.0$ is presented in Fig.\,\ref{Poincare-BMN1} (f).
It is computed by taking an initial condition randomly from the region around $r=0.0$, $p_r=3.8$, $x=-0.50$, $p_2=2.33$
with the radius $5.0\times 10^{-3}$. 
The largest Lyapunov exponent is approximately $0.302$\,, 
which is significantly non-zero. This result gives a support 
for the existence of chaos. 

\subsubsection*{(2) \quad  A single Coulomb potential $(N=3)$}

As the next example, let us consider the following $3 \times 3$ ansatz: 
\begin{align}
 &X^1=\sqrt{6}\,x(t)
 \begin{pmatrix}
  2/3&&\\
  &-1/3 & \\
  & & -1/3
 \end{pmatrix}\,,\quad
 X^2=\sqrt{6}\,x(t)
 \begin{pmatrix}
  -1/3&&\\
  &2/3 & \\
  & & -1/3
 \end{pmatrix}\,,\quad
 \nonumber \\
 &
 X^3=\sqrt{6}\,x(t)
 \begin{pmatrix}
  -1/3&&\\
  &-1/3 & \\
  & & 2/3
 \end{pmatrix}\,,\qquad \qquad 
 \nonumber \\
 &
 X^4=y(t)
 \begin{pmatrix}
  0&&\\
  & 0 &-i{\rm e}^{-i\varphi} \\
  & i {\rm e}^{i\varphi} &0
 \end{pmatrix}\,,\quad
 X^5=y(t)
 \begin{pmatrix}
  0&& i{\rm e}^{i\varphi}\\
  &0 & \\
  -i{\rm e}^{-i\varphi}& &0
 \end{pmatrix}\,,\quad
 \nonumber \\
 &
 X^6=y(t)
 \begin{pmatrix}
  0&-i{\rm e}^{-i\varphi}&\\
  i{\rm e}^{i\varphi}&0 & \\
  & &0
 \end{pmatrix}\,,\quad
 X^7=y(t)
 \begin{pmatrix}
  0&&\\
  &0 & {\rm e}^{-i\varphi} \\
  & {\rm e}^{i\varphi} &0
 \end{pmatrix}\,,\quad
 \nonumber \\
 &
 X^8=y(t)
 \begin{pmatrix}
  0&& {\rm e}^{i\varphi}\\
  &0 & \\ 
  {\rm e}^{-i\varphi}& &0
 \end{pmatrix}\,, \quad
 X^9=y(t)
 \begin{pmatrix}
  0& {\rm e}^{-i\varphi}&\\
  {\rm e}^{i\varphi}&0 & \\
  & &0
 \end{pmatrix}\,.
 \label{ansatz2}
\end{align}
Here $x(t)$ and $y(t)$ are real functions to be determined. The angle variable $\varphi(t)$ is cyclic 
and the associated constant of motion is $L\equiv y^2\dot \varphi$\,.
This ansatz (\ref{ansatz2}) was originally proposed by Arnlind and Hoppe \cite{Arnlind:2003nh}. 
Then the resulting differential equations are given by 
\begin{align}
 0&=
 \ddot x
 +\left( \frac{\mu}{3} \right)^2x
 +12xy^2,
 \nonumber \\
 0&=
 \ddot y
 +\left( \frac{\mu}{6} \right)^2y
 +8y^3
 +12x^2y
 -\frac{L^2}{y^3}\,. \label{s2}
\end{align}
The system (\ref{s2}) can also be obtained from the Lagrangian, 
\begin{align}
 &L=
 \frac{1}{2}\dot x^2
 -2\left( \frac{\mu}{6} \right)^2 x^2
 +\frac{1}{2}\dot y^2
 -\frac{1}{2}\left( \frac{\mu}{6} \right)^2 y^2
 -2y^4
 -\frac{L^2}{2y^2}
 -6x^2y^2.
\end{align}
This system can be seen as a system composed of two-coupled anharmonic oscillators 
with a repulsive Coulomb-type potential. 

\medskip 

\begin{figure}[htbp]
\begin{tabular}{cc}
\includegraphics[scale=.3,angle=-90]{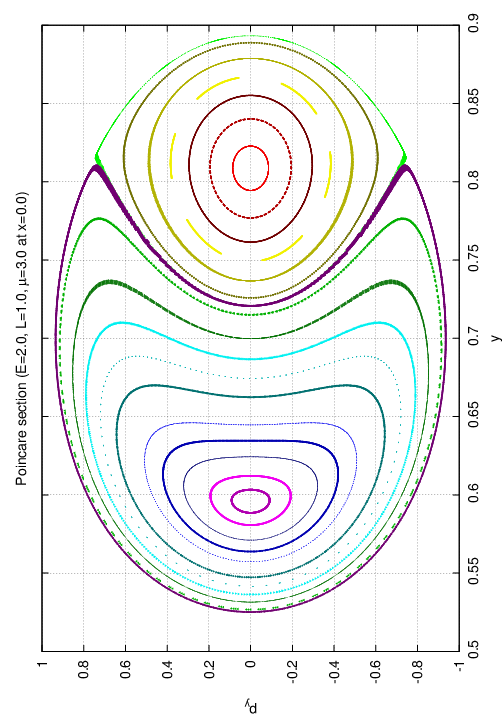} & 
\includegraphics[scale=.3,angle=-90]{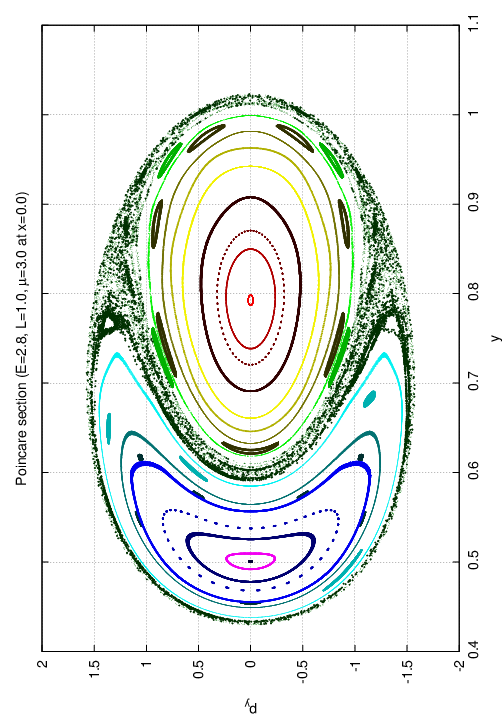} \\
{\footnotesize (a) \quad Poincar\'e section with $E=2.0$} & 
{\footnotesize (b) \quad Poincar\'e section with $E=2.8$} \\
\includegraphics[scale=.3,angle=-90]{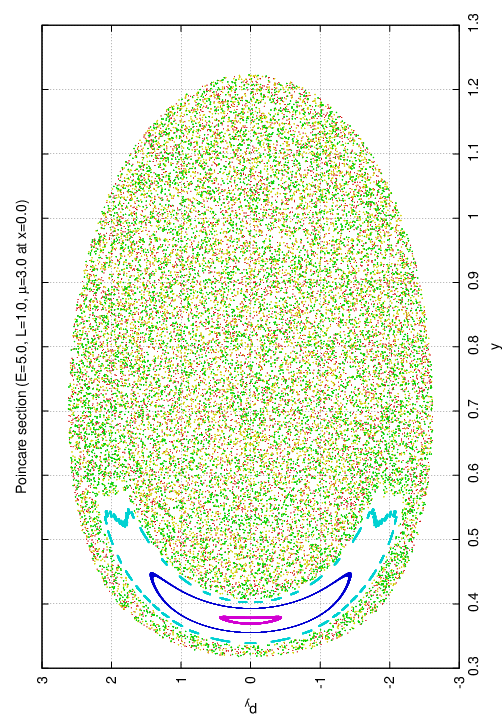} & 
\includegraphics[scale=.3,angle=-90]{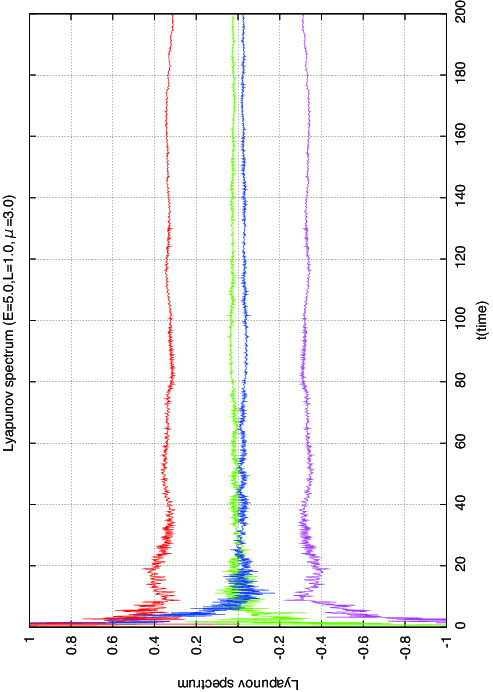} \\
{\footnotesize (c) \quad Poincar\'e section with $E=5.0$} & 
{\footnotesize (d) \quad Lyapunov spectrum with $E=5.0$} 
\end{tabular}
\caption{ \label{Poincare-BMN2}
\footnotesize Poincar\'e sections and a Lyapunov spectrum 
with the ansatz \eqref{ansatz2}. 
In Fig.\ (a), two KAM tori are contiguous.
In Fig.\ (b), local chaos appears around the boundary.
}
\end{figure}

Poincar\'e sections are shown in Figs.\,\ref{Poincare-BMN2} (a)-(c). 
The sections are taken with $x=0$ and $\dot{x}>0$\,. 
For $E=2.0$\,, 2.8 and 5.0 with $\mu=3.0$\,, the sections are numerically computed.  
When $E=2.0$\,, trajectories are (quasi-)periodic [Fig.\,\ref{Poincare-BMN2} (a)]\,. 
The left KAM tori are pushed by the right KAM tori, but the energy is not sufficiently high 
and hence there is no chaotic behavior.  
When $E=2.8$\,, the energy is high enough to destroy 
the contiguous KAM tori in Fig.\,\ref{Poincare-BMN2} (a). 
Hence chaotic motions appear along the contiguous line [Fig.\,\ref{Poincare-BMN2} (b)]. 
This behavior would be understood as local chaos again, according to the Chirikov 
overlap criterion.  
When $E=5.0$\,, almost trajectories become chaotic
and few ones remain (quasi-)periodic. 

\medskip

The Lyapunov spectrum in Fig.\,\ref{Poincare-BMN2} (d) also supports the existence of chaos. 
It is computed by taking an initial condition randomly from the region around $x=0.95$, $p_x=0.0$, $y=0.5$, $p_2=1.44$
with the radius $5.0\times 10^{-3}$\,. 

\subsubsection*{(3) \quad Integrable fuzzy spheres $(N=4)$}

Let us concentrate on the $SO(3)$ subsector and 
consider the following ansatz:
\begin{align}
 X^i= \frac{1}{2}
 \begin{pmatrix}
  r(t)\, \sigma^i & x(t)\, \sigma^i \\
  x(t)\, \sigma^i & r(t)\, \sigma^i
 \end{pmatrix}\,,\qquad
 X^a=0\,. 
 \label{ansatz3}
\end{align}
Here $\sigma^i$ are the standard Pauli matrices and hence the matrix size of $X^i$ is $4 \times 4$\,. 
Then $r(t)$ and $x(t)$ are real functions to be determined. The ansatz (\ref{ansatz3}) satisfies the Gauss 
law constraint (\ref{Gauss}). With the ansatz (\ref{ansatz3})\,, 
the equations of motion (\ref{eom}) are boiled down to  
\begin{align}
 0&=
 \ddot r
 +2r\left( r-\frac{\mu}{3} \right) \left( r-\frac{\mu}{6} \right)
 +6\left( r-\frac{\mu}{6} \right) x^2\,,
 \nonumber \\
 0&=
 \ddot x
 +2x^3
 +\left( \frac{\mu}{3} \right)^2 x
 +\left( -2\mu r+6r^2 \right) x\,. \label{s3}
\end{align}
One can derive this system (\ref{s3}) from the following Lagrangian:
\begin{align}
 L=
 \frac{1}{2}\dot r^2
 -\frac{1}{2}r^2\left( r-\frac{\mu}{3} \right)^2
 +\frac{1}{2}\dot x^2
 -\frac{1}{2}\left( \frac{\mu}{3} \right)^2 x^2
 -\frac{1}{2}x^4
 -3r\left( r-\frac{\mu}{3} \right)x^2\,.
\end{align}

\medskip 

Note that the system (\ref{s3}) is completely integrable. 
This means that there exists the second integral of 
motion as well as the Hamiltonian. In fact, the integrability of this system was shown in \cite{int}\,. 
In order to show the integrability, it is helpful to shift $r$ as $r = y + \mu/6$\,. 
Then the resulting Lagrangian is  
\begin{eqnarray}
L = \frac{1}{2}\dot{x}^2 + \frac{1}{2}\left(\frac{\mu}{6}\right)^2x^2 -\frac{1}{2}x^4 
+ \frac{1}{2}\dot{y}^2 + \frac{1}{2}\left(\frac{\mu}{6}\right)^2 y^2 -\frac{1}{2}y^4 
-3x^2y^2\,,
\label{ex}
\end{eqnarray}
where the constant term has been dropped off. 
Now it is easy to compare the expression (\ref{ex}) with the results in \cite{int}\,, 
and the second integral is given by
\begin{align}
I= -\dot{x} \dot{y} - 2xy \left[ x^2 + y^2 - \left( \frac{\mu}{6} \right)^2 \right]\,. 
\end{align}

\medskip 

It may be interesting to rewrite the Hamiltonian $H$ by subtracting $I$\,. 
One can easily show the following expression:
\begin{eqnarray}
H - I = \frac{1}{2}P^2 -\left(\frac{\mu}{6}\right)^2 R^2 +  \frac{1}{2}R^4 \equiv H_R\,. 
\end{eqnarray}
Here $R \equiv x+y$ and $P$ is the conjugate momentum for $R$\,. 
Thus the Hamiltonian $H= H_R + I$ is nothing but a system of two independent oscillators. 

\medskip 

Finally, it should be remarked that the following orthogonal transformation 
makes the integrability apparent: 
\begin{align}
 \frac{1}{\sqrt{2}}
 \begin{pmatrix}
  \mathbf{1}_2&\mathbf{1}_2\\
  \mathbf{1}_2&-\mathbf{1}_2
 \end{pmatrix}\,.
\end{align}
It recasts $X^i$ into the block-diagonal form that indicates that 
the system is composed of two independent pulsating fuzzy spheres 
with the radial coordinates $r+x$ and $r-x$\,, respectively.
Although this ansatz seems rather trivial, 
it would be worth to mention that the BMN matrix model contains 
an integrable subsystem.

\section{Conclusion and Discussion}

In this paper, we have studied classical chaotic motions in the BMN matrix model. 
We have examined three ans\"atze:   
1) two pulsating fuzzy spheres, 2) a single Coulomb-type potential, and 3) two coupled fuzzy spheres. 
For the first two cases, we have shown the existence of chaos 
by computing Poincar\'e sections and a Lyapunov spectrum. 
The third case has led to an integrable system. As a result, the BMN matrix model is not integrable 
in the sense of Liouville, though there may be some integrable subsectors. 

\medskip 

It would be worth to comment on preceding works on the integrability of the BMN matrix model 
\cite{KKP,Klose1,Klose2,Klose3,Klose4}. The integrability argument is based on the fact that 
the BMN matrix model is obtained as a consistent Kaluza-Klein (KK) reduction of the $\mathcal{N}$=4 
super Yang-Mills (SYM) theory  
on $R\times S^3$ \cite{KKP}. 
The dilatation operator of the $\mathcal{N}$=4 SYM on ${\mathbf R}^{1,3}$ 
is identified with the Hamiltonian of the $\mathcal{N}$=4 SYM on $R\times S^3$ 
via the operator-state mapping.  
It is now realized that the dilatation operator is closely connected with 
an integrable spin chain Hamiltonian \cite{MZ,BDS,BS}. 
Hence the integrability of the dilatation operator 
might be inherited to the BMN matrix model through the KK-reduction. 
Note here that we have considered classical chaos in the BMN matrix model, 
which is concerned with the classical Hamiltonian of the $\mathcal{N}$=4 SYM on $R\times S^3$ 
(or the classical dilatation operator $D_{\rm cl}$ of the $\mathcal{N}$=4 SYM on ${\mathbf R}^{1,3}$)\,. 
On the other hand, the spin-chain description is realized at the quantum level 
and the quantum dilatation operator $D_{\rm qu}$ plays a central role\footnote{More precisely, 
$D_{\rm qu}$ is divided as follows: 
\[
D_{\rm qu} = D_{\rm tree} + \sum_{n=1}^{\infty}D_{n-{\rm loop}}\,.
\]
The first term $D_{\rm tree}$ is concerned with the free Hamiltonian of the BMN matrix model, 
in which there is no chaotic motion as a matter of course. 
}\,. Thus our analysis has indirect relevance with the preceding one. 
It may be interesting to study whether the classical chaos can survive the quantization. 

\medskip 

The BMN matrix model was originally proposed in the context of M-theory 
and it can be regarded as a deformation of the M(atrix) theory \cite{BFSS}. 
It would be very intriguing to try to understand the chaotic motions 
from the point of view of the D0-brane dynamics. 
With some generalizations, those may capture an aspect of black hole physics  
such as the fast scrambling process \cite{SS}. For interesting works along this direction, 
see \cite{Berenstein}. In the very recent, a bound on a Lyapunov exponent has been argued \cite{MSS}. 

\medskip 

It seems promising that non-linear dynamical perspectives of string theory and M-theory 
would be more and more important. 
The way and result presented here would be a key ingredient 
in tackling the fundamental problems like information loss 
due to a black hole formation.

\subsection*{Acknowledgments}

It is a pleasure to acknowledge helpful discussions with Sinya Aoki, Masanori Hanada, 
Takeshi Matsumoto and Shin-ichi Sasa. 
The work of YA was supported by the Japan Society for the Promotion of Science (JSPS). 
This work is supported in part by the JSPS Japan-Hungary Research Cooperative Program.  
A part of the numerical calculations were carried out on SR16000 at YITP in Kyoto University.

\appendix 

\section*{Appendix}

\section{How to compute the Lyapunov spectrum}

In Section 2, we have explained the notion of the Lyapunov spectrum in an intuitive way. 
More practically, in order to numerically evaluate the Lyapunov spectrum, 
we have to follow a mathematical formulation. 
The basic method is based on the expansion rate of the volume \cite{SN,BGGM}. 
With this method, all of the $N$ Lyapunov exponents can be computed. 

\medskip 

Consider a system of the first differential equations described by
\begin{eqnarray}
  \dot{{\bf x}} = {\bf F}\left({\bf x}\right)\,, \qquad {\bf x} = (x^1,\cdots, x^i, \cdots, x^N)^t\,.
\end{eqnarray}
Then, by taking a variation, the following relation is obtained, 
\begin{eqnarray}
  \delta \dot{x}^i = \frac{\partial {F^i}}{\partial {x^j}}\, \delta {x^j}\,. 
  \label{eq:dev.dif.eq}
\end{eqnarray}
Hence a deviation vector $\delta \mathbf{x}\left(t\right)$ can be expressed as 
\begin{eqnarray}
  \delta \mathbf{x}\left(t\right) = U_t\, \delta \mathbf{x}\left(0\right)\,. 
\end{eqnarray}
Here $U_t$ is a time evolution operator and $\delta \mathbf{x}\left(0\right)$ 
is the initial value of the deviation vector. This is a solution of (\ref{eq:dev.dif.eq}).

\medskip 

The linearity of $U_t$ leads to the following relation: 
\begin{eqnarray}
  \delta \mathbf{x}\left(t+s\right) = U_t \delta \mathbf{x} \left(s\right)
  = U_t U_s \delta \mathbf{x} \left(0\right)\,.
\end{eqnarray}
Then the Lyapunov exponent for this deviation vector is defined as 
\begin{eqnarray}
  \label{eq:lyapunov.def.multi}
  \lambda \equiv \lim_{T\to\infty} \frac{1}{T} \log \frac{||\delta \mathbf{x}(T) ||}{||\delta \mathbf{x}(0) ||}
  = \lim_{n\to\infty} \frac{1}{n \Delta t} \log \frac{|| U_{\Delta t} \dots 
U_{\Delta t}\delta \mathbf{x}(0) ||}{||\delta \mathbf{x}(0) ||}\,.  
\end{eqnarray}
Here in the last expression $T$ is divided into $n$ small steps like $T = n\Delta t$\,. 
The convergence of this limit is proven by Oseledec \cite{Os}. 

\medskip 

\begin{figure}[htbp]
\begin{center}
\includegraphics[scale=.4,angle=0]{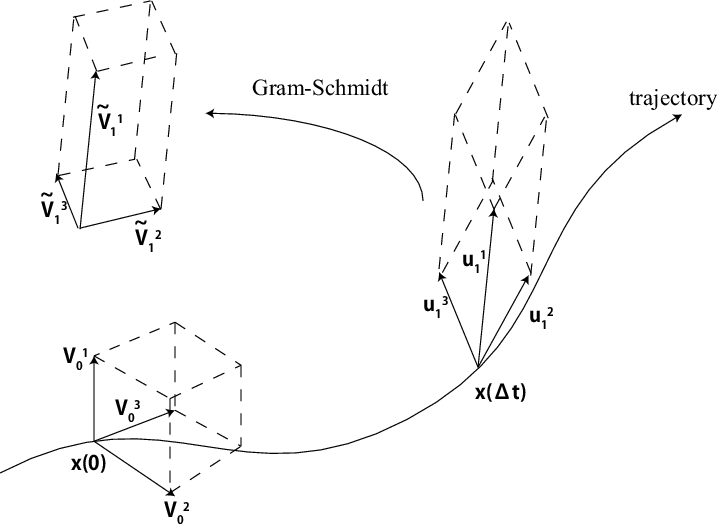}
\end{center}
\caption{ \label{fig:time.evolve}
\footnotesize Time evolution of a trajectory and a deviation vector set along it.
}
\end{figure}

Let us next introduce a fancy trick to compute the Lyapunov spectrum composed of $N$ exponents.
The first is to prepare an orthonormal basis set $\{\mathbf{v}_0^1, \mathbf{v}_0^2,\dots,\mathbf{v}_0^N\}$ 
to span the tangent vector space along the trajectory.
Then we make each basis $\mathbf{v}_0^i$ evolve by $U_{\Delta t}$\,. 
The resulting set of the vectors $\{\mathbf{u}_1^1, \mathbf{u}_1^2,\dots,\mathbf{u}_1^N\}$ 
cease to be one of the orthonormal vectors, 
where $\mathbf{u}_1^i = U_{\Delta t} \mathbf{v}_0^i$\,. This is depicted in Fig.\,\ref{fig:time.evolve}.

\medskip 

Then it is necessary to rebuild a new set of the orthogonal vectors 
$\{\mathbf{\widetilde{v}}_1^1, \mathbf{\widetilde{v}}_1^2, \dots, \mathbf{\widetilde{v}}_1^N\}$ 
from $\{\mathbf{u}_1^1, \mathbf{u}_1^2, \dots, \mathbf{u}_1^N\}$ by following the Gram-Schmidt process  
and compute the expansion rate $a_k$  for each vector $\mathbf{\widetilde{v}}_1^k$~: 
\begin{eqnarray}
  a_1^k \equiv \frac{||\mathbf{\widetilde{v}}_1^k||}{||\mathbf{v}_0^k||} = ||\mathbf{\widetilde{v}}_1^k||\,. 
\end{eqnarray}
Lastly, the orthogonal vector set $\{\mathbf{\widetilde{v}}_1^1, \mathbf{\widetilde{v}}_1^2, 
\dots, \mathbf{\widetilde{v}}_1^N\}$ is normalized as  
\begin{eqnarray}
  \mathbf{v}_1^k = \frac{1}{a_1^k} \mathbf{\widetilde{v}}_1^k  \qquad (k = 1,2,\dots,N)\,.
\end{eqnarray}
Thus a single cycle has been completed. Then the same cycle should be repeated $n$ times.
This means that the set $\{\mathbf{v}_i^1, \mathbf{v}_i^2, \dots, \mathbf{v}_i^N\}$ 
is used as an initial orthonormal vector set for the $(i+1)$-th cycle.

\medskip 

Now the largest Lyapunov exponent $\lambda_1$ is defined as 
\begin{eqnarray}
  \lambda_1 \equiv \lim_{n \to \infty}\, \frac{1}{n \Delta t}\, \sum_{i=1}^{n}\, \log a_i^1\,.
\end{eqnarray}
By construction, $\mathbf{v}_i^1$ heads toward the direction which has the highest sensitivity to initial conditions 
at sufficiently late time. Thus the expansion rate $a_i^1$ has the largest value in this region.
Then it is obvious from (\ref{eq:lyapunov.def.multi}) that the largest Lyapunov exponent is given 
by the sum of logarithm of the expansion rate. 

\medskip 

In a similar way, the $k$-th Lyapunov exponent $\lambda_k$ is defined as  
\begin{eqnarray}
  \lambda_k \equiv \lim_{n \to \infty} \,\frac{1}{n \Delta t}\, \sum_{i=1}^{n}\, \log a_i^k\,.  
\end{eqnarray}
Thus the set of the exponents $\{\lambda_1,\ldots, \lambda_N\}$ is obtained. 
This set is called the Lyapunov spectrum.  
Note that the computation time can be estimated by the inverse of 
the smallest positive Lyapunov exponent. This is because all of the exponents converge within this scale.

\medskip 

A characteristic of the Lyapunov spectrum in the Hamiltonian system is that 
the total sum of the exponents becomes zero, i.e., 
\begin{eqnarray}
\label{eq:lyap.consv}
\sum_{k=1}^N \lambda_k = 0\,.
\end{eqnarray}
This relation follows from the computation scheme of the Lyapunov spectrum.
By construction, each of the exponent measures the expansion rate of the orthogonal basis 
in the phase space. Hence the volume  $V(t)$ spanned by the bases is expressed as 
\begin{eqnarray}
V(t) = \exp\left(t\sum_{k=1}^N \lambda_k \right) V(0)\,. 
\end{eqnarray}
Liouville's theorem says that the phase-space volume is conserved in the Hamiltonian system (i.e., $V(t)=V(0)$). 
Thus the relation (\ref{eq:lyap.consv}) must be satisfied. 

\medskip 

As a side note, for the dissipative system, the following inequality is satisfied:
\begin{eqnarray}
\label{eq:lyap.disp}
\sum_{k=1}^N \lambda_k < 0\,. 
\end{eqnarray}
This is just because the volume $V(t)$ monotonically decreases.

\end{document}